\documentstyle[12pt]{article}
\advance \topmargin by -\headheight
\advance \topmargin by -\headsep

\evensidemargin \oddsidemargin
\begin{document}

\topmargin -.6in

\def\rf#1{(\ref{eq:#1})}
\def\lab#1{\label{eq:#1}}
\def\nonu{\nonumber}
\def\br{\begin{eqnarray}}
\def\er{\end{eqnarray}}
\def\be{\begin{equation}}
\def\ee{\end{equation}}
\def\eq{\!\!\!\! &=& \!\!\!\! }
\def\foot#1{\footnotemark\footnotetext{#1}}
\def\lb{\lbrack}
\def\rb{\rbrack}
\def\llangle{\left\langle}
\def\rrangle{\right\rangle}
\def\blangle{\Bigl\langle}
\def\brangle{\Bigr\rangle}
\def\llb{\left\lbrack}
\def\rrb{\right\rbrack}
\def\Blb{\Bigl\lbrack}
\def\Brb{\Bigr\rbrack}
\def\lcurl{\left\{}
\def\rcurl{\right\}}
\def\({\left(}
\def\){\right)}
\def\v{\vert}                     
\def\bv{\bigm\vert}               
\def\Bgv{\;\Bigg\vert}            
\def\bgv{\bigg\vert}              
\def\lskip{\vskip\baselineskip\vskip-\parskip\noindent}
\def\mskp{\par\vskip 0.3cm \par\noindent}
\def\sskp{\par\vskip 0.15cm \par\noindent}
\def\bc{\begin{center}}
\def\ec{\end{center}}
\def\Lbf#1{{\Large {\bf {#1}}}}
\def\lbf#1{{\large {\bf {#1}}}}
\relax


\def\mark{\noindent{\bf Remark.}\quad}
\def\prop{\noindent{\bf Proposition.}\quad}
\def\theor{\noindent{\bf Theorem.}\quad}
\def\name{\noindent{\bf Definition.}\quad}
\def\exam{\noindent{\bf Example.}\quad}
\def\proof{\noindent{\bf Proof.}\quad}

\def\tr{\mathop{\rm tr}}                  
\def\Tr{\mathop{\rm Tr}}                  
\newcommand\partder[2]{{{\partial {#1}}\over{\partial {#2}}}}
\newcommand\funcder[2]{{{\delta {#1}}\over{\delta {#2}}}}
\newcommand\Bil[2]{\Bigl\langle {#1} \Bigg\vert {#2} \Bigr\rangle}  
\newcommand\bil[2]{\left\langle {#1} \bigg\vert {#2} \right\rangle} 
\newcommand\me[2]{\langle {#1}\vert {#2} \rangle} 

\newcommand\sbr[2]{\left\lbrack\,{#1}\, ,\,{#2}\,\right\rbrack} 
\newcommand\Sbr[2]{\Bigl\lbrack\,{#1}\, ,\,{#2}\,\Bigr\rbrack} 
\newcommand\pbr[2]{\{\,{#1}\, ,\,{#2}\,\}}       
\newcommand\Pbr[2]{\Bigl\{ \,{#1}\, ,\,{#2}\,\Bigr\}}  
\newcommand\pbbr[2]{\lcurl\,{#1}\, ,\,{#2}\,\rcurl}  
\newcommand\sumi[1]{\sum_{#1}^{\infty}}   

\def\a{\alpha}
\def\b{\beta}
\def\c{\chi}
\def\d{\delta}
\def\D{\Delta}
\def\eps{\epsilon}
\def\vareps{\varepsilon}
\def\g{\gamma}
\def\G{\Gamma}
\def\grad{\nabla}
\def\h{{1\over 2}}
\def\l{\lambda}
\def\L{\Lambda}
\def\m{\mu}
\def\n{\nu}
\def\ov{\over}
\def\om{\omega}
\def\O{\Omega}
\def\p{\phi}
\def\P{\Phi}
\def\pa{\partial}
\def\tpa{{\tilde \partial}}
\def\pr{\prime}
\def\ra{\rightarrow}
\def\lra{\longrightarrow}
\def\s{\sigma}
\def\S{\Sigma}
\def\t{\tau}
\def\th{\theta}
\def\Th{\Theta}
\def\z{\zeta}
\def\ti{\tilde}
\def\wti{\widetilde}
\def\one{\hbox{{1}\kern-.25em\hbox{l}}}

\def\cA{{\cal A}}
\def\cB{{\cal B}}
\def\cC{{\cal C}}
\def\cD{{\cal D}}
\def\cE{{\cal E}}
\def\cH{{\cal H}}
\def\cL{{\cal L}}
\def\cM{{\cal M}}
\def\cN{{\cal N}}
\def\cP{{\cal P}}
\def\cQ{{\cal Q}}
\def\cR{{\cal R}}
\def\cS{{\cal S}}
\def\cU{{\cal U}}
\def\cV{{\cal V}}
\def\cW{{\cal W}}
\def\cY{{\cal Y}}

\def\phanta{\phantom{aaaaaaaaaaaaaaa}}
\def\phantb{\phantom{aaaaaaaaaaaaaaaaaaaaaaaaa}}
\def\phantc{\phantom{aaaaaaaaaaaaaaaaaaaaaaaaaaaaaaaaaaa}}

 \def\Winf{{\bf W_\infty}}               
\def\Win1{{\bf W_{1+\infty}}}           
\def\nWinf{{\bf {\hat W}_\infty}}       
\def\PsDA{\Psi{\cal DO}}
\def\Intres{\int dx\, {\rm Res} \; }

\def\KP3{{\bf KP_{2+1}}}
\def\mKP3{{\bf mKP_{2+1}}}
\def\KPm{{\bf (m)KP_{2+1}}}
\def\KPt{{\bf KP_{1+1}}}
\def\mKPt{{\bf mKP_{1+1}}}

\newcommand{\nit}{\noindent}
\newcommand{\ct}[1]{\cite{#1}}
\newcommand{\bi}[1]{\bibitem{#1}}
\newcommand\PRL[3]{{\sl Phys. Rev. Lett.} {\bf#1} (#2) #3}
\newcommand\NPB[3]{{\sl Nucl. Phys.} {\bf B#1} (#2) #3}
\newcommand\NPBFS[4]{{\sl Nucl. Phys.} {\bf B#2} [FS#1] (#3) #4}
\newcommand\CMP[3]{{\sl Commun. Math. Phys.} {\bf #1} (#2) #3}
\newcommand\PRD[3]{{\sl Phys. Rev.} {\bf D#1} (#2) #3}
\newcommand\PLA[3]{{\sl Phys. Lett.} {\bf #1A} (#2) #3}
\newcommand\PLB[3]{{\sl Phys. Lett.} {\bf #1B} (#2) #3}
\newcommand\JMP[3]{{\sl J. Math. Phys.} {\bf #1} (#2) #3}
\newcommand\PTP[3]{{\sl Prog. Theor. Phys.} {\bf #1} (#2) #3}
\newcommand\SPTP[3]{{\sl Suppl. Prog. Theor. Phys.} {\bf #1} (#2) #3}
\newcommand\AoP[3]{{\sl Ann. of Phys.} {\bf #1} (#2) #3}
\newcommand\RMP[3]{{\sl Rev. Mod. Phys.} {\bf #1} (#2) #3}
\newcommand\PR[3]{{\sl Phys. Reports} {\bf #1} (#2) #3}
\newcommand\FAP[3]{{\sl Funkt. Anal. Prilozheniya} {\bf #1} (#2) #3}
\newcommand\FAaIA[3]{{\sl Functional Analysis and Its Application} {\bf #1}
(#2) #3}
\def\InvM#1#2#3{{\sl Invent. Math.} {\bf #1} (#2) #3}
\newcommand\LMP[3]{{\sl Letters in Math. Phys.} {\bf #1} (#2) #3}
\newcommand\IJMPA[3]{{\sl Int. J. Mod. Phys.} {\bf A#1} (#2) #3}
\newcommand\TMP[3]{{\sl Theor. Mat. Phys.} {\bf #1} (#2) #3}
\newcommand\JPA[3]{{\sl J. Physics} {\bf A#1} (#2) #3}
\newcommand\JSM[3]{{\sl J. Soviet Math.} {\bf #1} (#2) #3}
\newcommand\MPLA[3]{{\sl Mod. Phys. Lett.} {\bf A#1} (#2) #3}
\newcommand\JETP[3]{{\sl Sov. Phys. JETP} {\bf #1} (#2) #3}
\newcommand\JETPL[3]{{\sl  Sov. Phys. JETP Lett.} {\bf #1} (#2) #3}
\newcommand\PHSA[3]{{\sl Physica} {\bf A#1} (#2) #3}
\newcommand\PHSD[3]{{\sl Physica} {\bf D#1} (#2) #3}

\newcommand\hepth[1]{{\sl hep-th/#1}}
\newcommand\Back{B\"{a}cklund}

\begin{titlepage}
\vspace*{-1cm}
\noindent
December, 1994 \hfill{BGU-94 / 25 / December - PH}\\
\phantom{bla}
\hfill{UICHEP-TH/94-12}\\
\phantom{bla}
\hfill{hep-th/9501018}
\\
\begin{center}
{\large {\bf Darboux-\Back ~Solutions of ${\bf SL(p,q)}$ KP-KdV Hierarchies, \\
Constrained Generalized Toda Lattices, \\
and Two-Matrix String Model}}
\end{center}
\vskip .3in
\begin{center}
{ H. Aratyn\footnotemark
\footnotetext{Work supported in part by the U.S. Department of Energy
under contract DE-FG02-84ER40173}}
\par \vskip .1in \noindent
Department of Physics \\
University of Illinois at Chicago\\
845 W. Taylor St.\\
Chicago, IL 60607-7059, {\em e-mail}:
aratyn@uic.edu \\
\par \vskip .3in
{ E. Nissimov$^{\,2}$  and S. Pacheva \foot{Supported in part by Bulgarian NSF
grant {\em Ph-401}.}}
\par \vskip .1in \noindent
Institute of Nuclear Research and Nuclear Energy \\
Boul. Tsarigradsko Chausee 72, BG-1784 ~Sofia, Bulgaria \\
{\em e-mail}: emil@bgearn.bitnet, svetlana@bgearn.bitnet \\
and \\
Department of Physics, Ben-Gurion University of the Negev \\
Box 653, IL-84105 $\;$Beer Sheva, Israel \\
{\em e-mail}: emil@bgumail.bgu.ac.il, svetlana@bgumail.bgu.ac.il
\end{center}
\vskip .3in

\begin{abstract} \noindent
We present an unifying description of the graded $SL(p,q)$ KP-KdV hierarchies,
including the Wronskian construction of their tau-functions as well as
the coefficients of the pertinent Lax operators, obtained via successive
applications of special Darboux-B\"{a}cklund transformations.
The emerging Darboux-B\"{a}cklund structure is identified as a constrained
generalized Toda lattice system relevant for the two-matrix string model.
It allows simple derivation of the $n$-soliton solutions of the unconstrained
KP system. Also, the exact Wronskian solution for the two-matrix model
partition function is found.
\end{abstract}

\end{titlepage}

\noindent
{\large {\bf 1. Introduction}}
\mskp

The Kadomtsev-Petviashvili (KP) hierarchy of integrable soliton nonlinear
evolution equations \ct{Zakh-Dickey} is among the most important physically
relevant integrable systems. One of the main reasons for the interest towards
the KP hierarchy in the last few years originates from its deep connection
with the statistical-mechanical models of random matrices
((multi-)matrix models) providing non-perturbative discretized formulation
of string theory \ct{integr-matrix}. Most of the
studies in the latter area follow the ideas of the original papers
\ct{double-scale}, where the integrability structures arise only after
taking the continuum {\em double-scaling} limit. There exists, however, an
alternative efficient approach \ct{BX} for extracting continuum differential
integrable hierarchies from multi-matrix string models even {\em before}
taking the continuum limit. More precisely, it is various reductions of the
full KP hierarchy ({\em constrained} KP Hierarchies) which play the major
r\^{o}le in the latter context.

On the other hand, constrained KP hierarchies arise also naturally
in purely solitonic context as shown below in subsection (6.2)
(see also \ct{Oevel-Dickey} and references therein).

It is the aim of the present note to study various properties and provide
exact solutions for a specific class of constrained KP hierarchies -- the
graded $SL(p,q)$ KP-KdV hierarchies, which are intimately related with
the two-matrix string model (which is the most physically relevant one).
The following main results are contained in the sequel:
\sskp
(1) We establish the equivalence between conventional ``symmetry''-constrained
KP hierarchies \ct{Oevel-Dickey} and multi-boson reductions of the full KP
hierarchy \ct{ANPV,BXL}, also known as graded $SL(p,q)$-type KP-KdV
hierarchies \ct{Yu,office}, which appear in two-matrix models of string theory
\ct{BX}. In particular, we provide the explicit Miura map relating the above
hierarchies.
\sskp
(2) Explicit exact solutions are found for $SL(p,q)$ KP-KdV integrable systems,
including eigenfunctions and $\t$-functions, via special Darboux-\Back (DB)
transformations.
\sskp
(3) We establish equivalence between the set of successive DB transformations
on the $SL(p,1)$ KP-KdV system and the equations of motion of a {\em
constrained} generalized Toda lattice model, which embodies the integrability
structure of two-matrix string models.
\sskp
(4) As a byproduct of (3) we obtain exact solution
for this constrained Toda lattice system under specific initial
conditions, relevant in the context of the two-matrix string model, and
derive the exact expression for the partition function of the latter.
\sskp
(5) The present DB formalism provides a simple systematic way to obtain the
$n$-soliton solutions for the full (unconstrained) KP system.
\lskip
{\large {\bf 2. Background on Generalized KP-KdV Hierarchies and Darboux-\Back
{}~Transformations}}
\mskp
{\bf 2.1 Constrained KP Hierarchies}

We shall consider the general class of constrained KP Lax operators
with higher purely-differential part \ct{Oevel-Dickey}, also known as
$N$-generalized two-boson KP Lax operators \ct{affine} \foot{In order
to avoid confusion, $D$ will denote differential operator in the sense of
pseudo-differential calculus, whereas derivative of a function will be denoted
as $\pa_x f$.} :
\br
L &=& L_{+} + \sum_{i=1}^N \Phi_i D^{-1} \Psi_i
\equiv D^r + \sum_{l=0}^{r-2} u_l D^l + \sum_{i=1}^N a_i \( D - b_i \)^{-1}
\lab{f-5}  \\
L_{+} &\equiv& D^r + \sum_{l=0}^{r-2} u_l D^l \qquad ; \quad
\Phi_i = a_i e^{\int b_i} \quad ,\quad \Psi_i = e^{-\int b_i} \lab{f-5a}
\er
One can also define an alternative consistent Poisson reduction of the
standard KP hierarchy based on the pseudo-differential Lax operators
of the form \ct{BX,ANPV,BXL} :
\br
L_N = L_{+} + \sum_{i=1}^N  r_i \prod_{k=i}^N D^{-1} q_k
= L_{+} + \sum_{i=1}^N A^{(N)}_i \prod_{k=i}^N \( D - B^{(N)}_k \)^{-1}
\phanta \lab{iss-8aaa} \\
r_i = A_{i} e^{\int B_{i}} \quad ;\;\; q_N = e^{-\int B_N} \;\; ,
\;\; q_j = e^{\int \( B_{j} - B_{j+1}\)} \;\;, \;\;
j =1,\ldots ,N-1  \lab{4.5}
\er
which we will call multi-boson reduction of the full KP Lax operator.
The above multi-boson reductions of the full KP Lax operators \rf{iss-8aaa}
are defining the generalized graded $SL(r+N,N)$ KP-KdV hierarchies
pertinent to the string two-matrix models (cf. refs.\ct{enjoy,office}).

In \ct{affine} these two formulations of the constrained KP hierarchy
have been related via successive DB similarity transformations.
Below in section 3 we will establish their complete equivalence
showing how the pseudo-differential Lax operators
\rf{f-5} and \rf{iss-8aaa} can be rewritten into each other
via a generalized Miura transformation.
Due to this result we can limit ourselves to study DB transformations
within the framework of the constrained KP hierarchy defined
by $N$-generalized two-boson KP Lax operators as in \rf{f-5}
\mskp
{\bf 2.2 On the DB transformations of the $N$-generalized two-boson KP Lax
operators}

The general form of a DB transformation on the
$N$-generalized two-boson KP Lax operator \rf{f-5} reads \ct{Oevel1,Chau} :
\br
{\wti L} &=& \chi D \chi^{-1} \( L_{+} + \sum_{i=1}^{N} \Phi_i D^{-1} \Psi_i \)
\chi D^{-1} \chi^{-1} \equiv {\wti L}_{+} + {\wti L}_{-}  \lab{baker-1} \\
{\wti L}_{+} &=& {L}_{+} + \chi \( \pa_x \( \chi^{-1} L_{+} \chi \)_{\geq 1}
D^{-1}\) \chi^{-1}    \lab{baker-2}  \\
{\wti L}_{-} &=& {\wti \Phi}_0 D^{-1} {\wti \Psi}_0 +
\sum_{i=1}^{N} {\wti \Phi}_i D^{-1} {\wti \Psi}_i   \lab{baker-3} \\
{\wti \Phi}_0 &=& \chi \llb \pa_x \( \chi^{-1} L_{+} \chi \) + \sum_{i=1}^N
\( \pa_x \( \chi^{-1}\Phi_i\) \pa_x^{-1} \( \Psi_i \chi\)
+ \Phi_i \Psi_i \) \rrb \equiv \( \chi D \chi^{-1} L \) \chi  \lab{baker-4} \\
{\wti \Psi}_0 &=& \chi^{-1} \qquad ,\quad
{\wti \Phi}_i = \chi \pa_x \( \chi^{-1} \Phi_i \) \qquad ,\quad
{\wti \Psi}_i = - \chi^{-1} \pa_x^{-1} \( \Psi_i \chi \)   \lab{baker-5}
\er
where all functions involved are (adjoint) eigenfunctions
of $L$ \rf{f-5}, {\sl i.e.}, they satisfy:
\be
\partder{}{t_n} f = L^{{n\over r}}_{+} f
\qquad f = \chi ,\Phi_i \quad ; \quad
\partder{}{t_n} \Psi_i = - {L^\ast}^{{n\over r}}_{+} \Psi_i
\lab{baker-6}
\ee
Let us particularly stress that the above eigenfunctions {\em are not}
Baker-Akhiezer eigenfunctions of $L$ from \rf{f-5}, unlike the construction
in ref.\ct{Oevel1}.

We are interested in the special case when $\chi$ coincides with one of the
original eigenfunctions of $L$, {\sl e.g.}  $\chi = \Phi_1$. Then
${\wti \Phi}_1 =0$ and the DB transformation \rf{baker-1} preserves the
$N$-generalized two-boson form \rf{f-5} of the Lax operators involved,
{\sl i.e.}, it becomes {\em auto}-\Back ~transformation.
Applying successive DB transformations in this case yields:
\br
L^{(k)} \eq T^{(k-1)} L^{(k-1)} \( T^{(k-1)}\)^{-1} =
\( L^{(k)}\)_{+} + \sum_{i=1}^N \Phi_i^{(k)} D^{-1} \Psi_i^{(k)} \quad ,\quad
T^{(k)} \equiv \Phi_1^{(k)} D \( \Phi_1^{(k)}\)^{-1}  \phantom{aaa}
\lab{shabes-3} \\
\Phi_1^{(k+1)} \eq \( T^{(k)} L^{(k)}\) \Phi_1^{(k)} \quad ,\quad
\Psi_1^{(k+1)} = \( \Phi_1^{(k)}\)^{-1}
\qquad, \qquad k =0,1,\ldots \lab{shabes-2} \\
\Phi_i^{(k+1)} \eq T^{(k)} \Phi_i^{(k)} \equiv
\Phi_1^{(k)} \pa_x \( \( \Phi_1^{(k)}\)^{-1} \Phi_i^{(k)}\)  \lab{shabes-5} \\
\Psi_i^{(k+1)} \eq - \( \Phi_1^{(k)}\)^{-1} \pa_x^{-1}
\( \Psi_i^{(k)}\Phi_1^{(k)} \) \qquad , \;\; i=2,\ldots , N  \lab{shabes-5a}
\er
Using the first identity \rf{shabes-3}, {\sl i.e.}, $ L^{(k+1)} T^{(k)}=
T^{(k)} L^{(k)}$ , one can rewrite \rf{shabes-2} in the form:
\be
\Phi_1^{(k)} = T^{(k-1)} T^{(k-2)} \cdots T^{(0)}
\( \( L^{(0)}\)^k \Phi_1^{(0)}\)
\lab{shabes-4}
\ee
whereas:
\be
\Phi_i^{(k)} = T^{(k-1)} T^{(k-2)} \cdots T^{(0)} \Phi_i^{(0)} \qquad ,\;\;
\quad i=2,\ldots , N  \lab{shabes-6}
\ee
Finally, for the coefficient of the next-to-leading differential term in
\rf{f-5} ~$ u_{r-2} = r \, Res\, L^{{1\over r}} = r\, \pa_x^2 \ln \t$, we
easily obtain from \rf{baker-2} (with $\chi =\Phi_1$) its $k$-step
DB-transformed expression:
\be
{1\over r} \( u_{r-2}^{(k)} - u_{r-2}^{(0)}\) =
\pa_x^2 \ln \frac{\t^{(k)}}{\t^{(0)}} =
\pa_x^2 \ln \( \Phi_1^{(k-1)} \cdots \Phi_1^{(0)}\)   \lab{sol-3-a}
\ee
\mskp
{\bf 2.3 Wronskian Preliminaries}
\sskp
Firstly we list three basic properties of Wronskian determinants.
\mskp
{\bf 1)} The derivative $\cD^{\pr}$ of a determinant $\cD$
of order $n$, whose entries are differentiable functions, can be written as:
\be
\cD^{\pr} = \cD_{(1)} + \cD_{(2)} + \ldots + \cD_{(n)}
\lab{derdet}
\ee
where $ \cD_{(i)} $ is obtained from $D$ by differentiating the entries in the
$i$-th row.
\mskp
{\bf 2)} {\sl Jacobi expansion theorem}:
\be
W_{k} \(f\) W_{k-1 }= W_{k } W_{k-1 }^{\pr} \(f\)- W_{k }^{\pr} W_{k-1 } \(f\)
\;\;{\rm or}\;\; \pa \( { W_{k-1 } (f) \over W_{k}} \) = {W_{k} \(f\) W_{k-1 }
\over W_{k}^2}
\lab{jac}
\ee
where the Wronskians are $W_k \equiv W_k \lb \psi_1, \ldots ,\psi_k \rb$
and $W_{k-1} \(f \)\equiv W_{k} \lb \psi_1, \ldots ,\psi_{k-1}, f\rb$.
For proof see \ct{am-AvM}.
Take a special class of Wronskians $W_n \equiv W_n \lb \psi, \psi^{\pr},
\ldots ,\pa^{n-1} \psi \rb$ .
Hence, from \rf{jac} we get :
\be
W_n W_n^{\pr \pr} - W_n^{\pr} W_n^{\pr} =
W_n \, W_{n-1}^{\pr}\(\pa^{n} \psi \) - W_n^{\pr}\, W_{n-1} \(\pa^{n} \psi \)
= W_{n-1} W_{n+1} \to \pa^2 \ln W_n = {W_{n+1} W_{n-1 } \over W_{n}^2}
\lab{jac-a}
\ee
\mskp
{\bf 3)} Iterative composition of Wronskians:
\be
T_k \, T_{k-1}\, \cdots\, T_1 (f ) \; =\; { W_{k} (f) \over W_k}
\lab{iw}
\ee
where
\be
T_j = { W_{j} \over W_{j-1} } D { W_{j-1} \over W_{j} } =
\( D + \( \ln { W_{j-1} \over W_{j} } \)^{\pr} \) \quad;\quad W_{0}=1
\lab{transf}
\ee
The proof of \rf{iw} follows by simple iteration of \rf{jac}
(see also the standard references on this subject
\ct{crum,ince,am-AvM}). For future use let us rewrite \rf{iw} as:
\be
\( D + v_k \) \( D + v_{k-1} \) \ldots \( D + v_1 \) \, f =
{ W_{k} (f) \over W_k} \quad;\qquad v_j \equiv \pa_x \ln { W_{j-1} \over W_j}
\lab{kw}
\ee
\lskip
{\bf 2.4 DB Solutions of Two-Bose KP System and Connection with Ordinary Toda
Lattice}
\sskp
The two-boson KP system defined by the Lax operator
$L = D + \Phi D^{-1} \Psi \equiv D + a \( D - b\)^{-1}$
is the most basic constrained KP structure.
We start with the initial ``free'' Lax operator $L^{(0)} = D$ and perform a DB
transformation:
\be
L^{(1)} = \(\Phi^{(0)} D {\Phi^{(0)}}^{-1}\)  \; D \;
\(\Phi^{(0)} D^{-1} {\Phi^{(0)}}^{-1}\)  
\, =\, D+ \llb \Phi^{(0)} \( \ln \Phi^{(0)}\)^{\pr \pr} \rrb
D^{-1} \(\Phi^{(0)}\)^{-1}    \lab{lone}
\ee
The construction below is a special application of property 3) and eq. \rf{iw}.

In the case under consideration the relevant formulas for successive DB
transformations \rf{shabes-2},\rf{shabes-3} specialize to:
\br
L^{(k+1)} \eq \(\Phi^{(k)}  D {\Phi^{(k)} }^{-1}\)  \; L^{(k)}
 \; \(\Phi^{(k)}  D^{-1} {\Phi^{(k)} }^{-1}\)
= D + \Phi^{(k+1)} D^{-1} \Psi^{(k+1)}
\lab{lkplus} \\
\Phi^{(k+1)} \eq \Phi^{(k)} \( \ln \Phi^{(k)}\)^{\pr \pr} + \(\Phi^{(k)}\)^2
\Psi^{(k)} \quad ,\quad \Psi^{(k+1)} = \(\Phi^{(k)}\)^{-1} \lab{pkplus}
\er
whereas eq.\rf{shabes-4} acquires the form (proved easily by induction in
$k$) :
\be
\Phi^{(k)} = { W_{k+1} \lb \P , \pa \P, \ldots , \pa^k \P \rb \over
W_k \lb \P , \pa \P, \ldots , \pa^{k-1} \P \rb}
\quad{\rm with}\quad \P = \P^{(0)}
\lab{phik}
\ee
Introduce now
\be
\p_k =\ln \Phi^{(k)} \quad \to \quad  \Phi^{(k)} = e^{\p_k}  \qquad k=0,\ldots
\lab{smallp}
\ee
which allows us to rewrite \rf{pkplus} as a (ordinary one-dimensional)
Toda lattice equation:
\be
\pa^2 \p_{k}\, =\, e^{\p_{k+1} - \p_{k}} - e^{\p_{k} - \p_{k-1}}
\lab{toda}
\ee
Introduce now new objects $\psi_n$ as follows:
\be
\p_n = \psi_{n+1} - \psi_{n} \quad \to \quad
\psi_{n} = \ln W_n \lb \P , \pa \P, \ldots , \pa^{n-1} \P \rb
\lab{psieqs}
\ee
{}From equation \rf{jac-a} we find immediately an equation for $\psi_n$:
\be
\pa^2 \psi_{n}= \pa^2 \ln W_n \, =\, e^{\psi_{n+1}+\psi_{n-1}-2\psi_{n}  }
\lab{newtoda}
\ee
with $\psi_n=0$ for $n \leq 0$.
We recognize at the right hand side of \rf{newtoda} a structure of
the Cartan matrix for $A_n$.
Leznov considered such an equation with Wronskian solution (in two dimensions)
in \ct{le80}.

Hence, the solutions of the (ordinary one-dimensional) Toda lattice equations,
with boundary conditions $\psi_n=0$ for $n \leq 0$, reproduce the DB
solutions of ordinary two-boson KP hierarchy \rf{phik}
upon taking into account that $ \Phi= \Phi^{(0)} = \exp \( \p_0\) =
\exp \( \psi_1 \) $.
\lskip
{\large {\bf 3. Equivalence between ${\bf N}$-Generalized Two-Boson and
${\bf 2N}$-Boson KP Hierarchies}}
\mskp
First, let us consider the simplest nontrivial case $N=2$ in \rf{f-5}.
Applying the simple identity:
\br
\phi D^{-1} \psi &=& \phi \psi \( \chi^{-1} D^{-1} \chi \) -
\phi D^{-1} \llb \chi \(\pa_x \frac{\psi}{\chi}\)\rrb
\( \chi^{-1}D^{-1}\chi\) \phanta   \nonu \\
&=& \phi \psi \( \chi^{-1} D^{-1} \chi \) -
\phi \frac{W\llb \chi ,\psi\rrb}{W\llb \chi\rrb}
\( \frac{W\llb \chi\rrb}{W\llb \chi ,\psi\rrb} D^{-1}
\frac{W\llb \chi ,\psi\rrb}{W\llb \chi\rrb}\) \( \chi^{-1}D^{-1}\chi\)
\lab{iss-4}
\er
for arbitrary functions $\phi,\psi,\chi$,
where in the second equality Wronskian identity \rf{iw} was used, we obtain:
\br
L &=& L_{+} + \Phi_1 D^{-1} \Psi_1 + \Phi_2 D^{-1} \Psi_2  \nonu \\
&=&L_{+} + A^{(2)}_2 \( D - B^{(2)}_2 \)^{-1} +
A^{(2)}_1 \( D - B^{(2)}_1 \)^{-1} \( D - B^{(2)}_2 \)^{-1}  \lab{iss-1} \\
A^{(2)}_2 &=& \Phi_1 \Psi_1 + \Phi_2 \Psi_2  \quad , \quad
B^{(2)}_2 = - \pa_x \ln \Psi_2  \lab{iss-2}  \\
A^{(2)}_1 &=& -\Phi_1 \frac{W\llb \Psi_2 ,\Psi_1\rrb}{W\llb \Psi_2\rrb}
\quad ,\quad
B^{(2)}_1 = -\pa_x \ln \frac{W\llb \Psi_2 ,\Psi_1\rrb}{W\llb \Psi_2\rrb}
\lab{iss-3}
\er

Using successively the same type of identity as \rf{iss-4}, together with
\rf{iw}, {\sl e.g.}, for arbitrary functions $\phi,\psi,\chi,\om$ :
\br
\phi D^{-1} \psi = \phi \psi \( \chi^{-1} D^{-1} \chi \) -
\phi \frac{W\llb \chi ,\psi\rrb}{W\llb \chi\rrb}
\( \frac{W\llb \chi\rrb}{W\llb \chi ,\om\rrb} D^{-1}
\frac{W\llb \chi ,\om\rrb}{W\llb \chi\rrb}\) \( \chi^{-1}D^{-1}\chi\) \nonu \\
+ \phi \frac{W\llb \chi ,\om ,\psi\rrb}{W\llb \chi ,\om \rrb}
\( \frac{W\llb \chi ,\om \rrb}{W\llb \chi ,\om ,\psi\rrb} D^{-1}
\frac{W\llb \chi ,\om\rrb}{W\llb \chi\rrb}\)
\( \frac{W\llb \chi\rrb}{W\llb \chi ,\om\rrb} D^{-1}
\frac{W\llb \chi ,\om\rrb}{W\llb \chi\rrb}\) \( \chi^{-1}D^{-1}\chi\)
\lab{iss-4a}
\er
we can prove by
induction in $N$ that the $N$-generalized two-boson KP Lax operator can be
transformed into $2N$-boson KP Lax operator:
\br
L &=& D^r + \sum_{l=0}^{r-2} u_l D^l + \sum_{i=1}^N \Phi_i D^{-1}\Psi_i
\equiv L_{+} + \sum_{i=1}^N a_i \( D - b_i \)^{-1}     \lab{iss-8aa}  \\
&=& L_{+} + \sum_{i=1}^N A^{(N)}_i \( D - B^{(N)}_i \)^{-1}
\( D - B^{(N)}_{i+1}\)^{-1} \cdots \( D - B^{(N)}_N \)^{-1}   \lab{iss-8a}
\er
upon the following change of variables, {\sl i.e.}, generalized Miura
transformation:
\br
A^{(N)}_k &=& (-1)^{N-k} \sum_{s=1}^k \Phi_s \frac{W\llb \Psi_N ,\ldots ,
\Psi_{k+1},\Psi_s \rrb}{W\llb \Psi_N ,\ldots ,\Psi_{k+1}\rrb} \lab{iss-8b}\\
B^{(N)}_k &=& - \pa_x \ln \frac{W\llb \Psi_N ,\ldots ,
\Psi_{k+1},\Psi_k \rrb}{W\llb \Psi_N ,\ldots ,\Psi_{k+1}\rrb} \lab{iss-8c}
\er

Let us now illustrate the equivalence between \rf{iss-8aa} and \rf{iss-8a} in
the inverse direction. To this end it is more convenient to use the $(r,q)$
form of \rf{iss-8aaa}.
Let us define the quantity
\be
Q_{k,i} \equiv (-1)^{i-k} \int q_i \int q_{i-1} \int \ldots
\int q_k \, (dx^{\pr})^{i-k+1}
\quad \qquad 1 \leq k \leq i \leq N
\lab{qni}
\ee
Then using that $ D^{-1} Q_{1,i-1}  q_i = D^{-1} Q_{1,i} D-
Q_{1,i}$ we obtain from the first eq.\rf{iss-8aaa}
\be
L_N = L_{+} + \sum_{i=2}^{N}  r_i^{(1)} \prod_{k=i}^N D^{-1} q_k
+ r_1 D^{-1} \( - Q_{1,N-1} q_N\)
\lab{rqlax-a}
\ee
where
\be
r_i^{(1)} \equiv r_i +r_1 Q_{1,i-1} \qquad i=2, \ldots , N
\lab{ronei}
\ee
The above process can be continued to yield expression \rf{f-5} with
\br
\P_i \eq r_i + \sum_{k=1}^{i-1} r_{k}
\sum_{s_{i-k-1} =s_{i-k-2} +1}^{i-k} \cdots \sum_{s_2 =s_{1} +1}^{i-k}
\sum_{s_{1} = 1}^{i-k} Q_{k,i-s_{i-k-1}-1}
Q_{i-s_{i-k-1},i-s_{i-k-2}-1} \cdots                    \nonu\\
&\cdots& Q_{i-s_2,i-s_{1}-1} Q_{i-s_{1},i-1}
\qquad \qquad\qquad \qquad \qquad 1 \leq i \leq N
\lab{piri} \\
\Psi_N \eq q_N \quad\; , \quad \;
\Psi_i = (-1)^{N-i} q_N \int q_{N-1} \int \ldots \int q_i
\, (dx^{\pr})^{N-i}  \qquad 1\leq i \leq N-1
\lab{psiqi}
\er
{\large {\bf 4. Exact Solutions of ${\bf SL(p,q)}$ KP-KdV via DB
Transformations}}
\mskp
Armed with the Wronskian identities from section 2.3, we can now represent the
$k$-step DB transformation \rf{shabes-4}---\rf{sol-3-a} in terms of
Wronskian determinants involving the coefficient functions of the ``initial''
Lax operator
\be
L^{(0)} = D^r + \sum_{l=0}^{r-2} u_l^{(0)} D^l +
\sum_{i=1}^N \Phi_i^{(0)} D^{-1} \Psi_i^{(0)}    \lab{seq-b}
\ee
only.
Indeed, using identity \rf{iw} and defining:
\be
\(L^{(0)}\)^k \P_1^{(0)} \equiv \chi^{(k)} \qquad \; k=1,2,\ldots
\lab{defchi}
\ee
we arrive at the following general result:

\prop
{\em The $k$-step DB-transformed eigenfunctions and the tau-function
\rf{shabes-4}---\rf{sol-3-a} of the $SL(r+N,N)$ KP-KdV system \rf{iss-8aa} for
arbitrary initial $L^{(0)}$ \rf{seq-b} are given by:
\br
\P_1^{(k)}&=& \frac{W_{k+1}\lb \P_1^{(0)},\chi^{(1)},\ldots, \chi^{(k)}\rb}
{W_{k}\lb \P_1^{(0)}, \chi^{(1)},\ldots ,\chi^{(k-1)}\rb}  \lab{pchi-a}  \\
\P_j^{(k)}&=&\frac{W_{k+1} \lb \P_1^{(0)},\chi^{(1)},\ldots ,\chi^{(k-1)},
\P_j^{(0)}\rb}{W_{k}\lb \P_1^{(0)}, \chi^{(1)}, \ldots,  \chi^{(k-1)}\rb}
\qquad , \;\; j=2,\ldots ,N    \lab{pchi-aa}  \\
\tau^{(k)} &=& W_{k} \lb  \P_1^{(0)}, \chi^{(1)},
\ldots,  \chi^{(k-1)}\rb \tau^{(0)}      \lab{tauok}
\er
where $\tau^{(0)}, \tau^{(k)}$ are the $\tau$-functions of $L^{(0)},
L^{(k)}$ , respectively, and $\chi^{(i)}$ is given by \rf{defchi}.}
\mskp

As an example let us consider the $SL(3,1)$ KP-KdV Lax operator, {\sl i.e.},
$r=2, N=1$ in \rf{f-5} (the latter is
pertinent to the simplest nontrivial string two-matrix model \ct{enjoy}) :
\be
L = D^2 + u + A \( D - B\)^{-1} = D^2 + u + \Phi D^{-1} \Psi \lab{sl-31}
\ee
{}From the basic formulas for successive DB transformations
\rf{shabes-2}--\rf{shabes-3}, applied to \rf{sl-31}, we have:
\br
L^{(k)} \eq D^2 + u^{(k)} + \Phi^{(k)} D^{-1} \Psi^{(k)}  \lab{sol-1} \\
L^{(k)} \!\!&\to&\!\!  L^{(k+1)} = T^{(k)} L^{(k)} \( T^{(k)}\)^{-1}
\quad ,\quad  T^{(k)} = \Phi^{(k)} D \( \Phi^{(k)}\)^{-1} \lab{sol-2} \\
u^{(k)} \eq 2 Res\, L^{\h} \equiv 2 \pa_x^2 \ln \t^{(k)} =
2 \pa_x^2 \ln \( \Phi^{(k-1)} \cdots \Phi^{(0)}\)  \lab{sol-3} \\
\Phi^{(k)} \!\!\!&\equiv& \!\!\!A^{(k)} e^{ \int B^{(k)}} =
\Phi^{(k-1)} \llb \pa_x \( {1\over {\Phi^{(k-1)}}} \pa_x^2
\Phi^{(k-1)} + 2 \pa_x^2 \ln \( \Phi^{(k-2)} \cdots \Phi^{(0)}\) \) +
\frac{\Phi^{(k-1)}}{\Phi^{(k-2)}} \rrb  \lab{sol-4} \\
\Psi^{(k)} \!\!&\equiv& \!\!
e^{ -\int B^{(k)}} = \( \Phi^{(k-1)}\)^{-1} \lab{sol-5} \\
u^{(0)} \eq 0 \quad ,\quad \Psi^{(0)} = 0 \quad ;\quad
\Phi^{(0)} = \int_{\Gamma} \frac{d\l}{2\pi} c(\l ) e^{\xi \( \l ,\{ t\}\)}
\quad ,\;\;
\xi \( \l ,\{ t\}\) \equiv \l x + \sum_{j \geq 2} \l^{j} t_j \lab{sol-6}
\er
where $\Phi^{(0)}$ in \rf{sol-6} is an arbitrary eigenfunction of the ``free''
$L^{(0)}= D^2$ (the contour $\Gamma$ in the complex $\l$-plane is chosen such
that the generalized Laplace transform of $c(\l )$ is well-defined).

As a corollary from the above proposition, we get in the case of \rf{sol-1} :
\br
\Phi^{(k)}& =& T^{(k-1)} \cdots T^{(0)} \( \pa_x^{2k} \Phi^{(0)} \) =
\frac{W \llb \Phi^{(0)}, \pa_x^2 \Phi^{(0)},
\ldots , \pa_x^{2k}\Phi^{(0)}\rrb}{ W \llb \Phi^{(0)}, \pa_x^2 \Phi^{(0)},
\ldots , \pa_x^{2(k-1)}\Phi^{(0)} \rrb} \lab{sol-7a} \\
\t^{(k)}& =& W \llb \Phi^{(0)}, \pa_x^2 \Phi^{(0)},
\ldots , \pa_x^{2(k-1)}\Phi^{(0)} \rrb  \lab{sol-8}
\er

Substituting \rf{sol-7a},\rf{sol-8} into \rf{sol-3}--\rf{sol-5} we obtain
the following explicit solutions for the coefficient functions of \rf{sl-31} :
\br
u^{(n)} &= &2 \pa_x^2 \ln W \llb \Phi^{(0)}, \pa_x^2 \Phi^{(0)},
\ldots , \pa_x^{2(n-1)}\Phi^{(0)} \rrb \phantom{aa} \lab{sol-3-aa} \\
B^{(n)} &=& \pa_x \ln \( \frac{W \llb \Phi^{(0)},
\pa_x^2 \Phi^{(0)},\ldots , \pa_x^{2(n-1)}\Phi^{(0)} \rrb}{W \llb
\Phi^{(0)}, \pa_x^2 \Phi^{(0)},\ldots , \pa_x^{2(n-2)}\Phi^{(0)} \rrb}\)
\lab{sol-3-b} \\
A^{(n)}&=&\frac{W \llb \Phi^{(0)},
\pa_x^2 \Phi^{(0)},\ldots , \pa_x^{2n}\Phi^{(0)} \rrb \, W \llb
\Phi^{(0)}, \pa_x^2 \Phi^{(0)},\ldots , \pa_x^{2(n-2)}\Phi^{(0)} \rrb}{\( W
\llb \Phi^{(0)},\pa_x^2 \Phi^{(0)},\ldots , \pa_x^{2(n-1)}\Phi^{(0)} \rrb\)^2}
\lab{sol-4-b}
\er

Similarly, in the more general case of
$SL(r+1,1)$ KP-KdV Lax operator for arbitrary finite $r$ :
\be
L = D^r + \sum_{l=0}^{r-2} u_l D^l + \Phi D^{-1} \Psi  \lab{sol-9}
\ee
which defines the integrable hierarchy corresponding to the general string
two-matrix model (cf. \ct{enjoy,office}), the generalizations of \rf{sol-7a}
and \rf{sol-8} read:
\br
\Phi^{(k)} \eq T^{(k-1)} \cdots T^{(0)} \( \pa_x^{k\cdot r} \Phi^{(0)} \)
= \frac{W \llb \Phi^{(0)}, \pa_x^{r} \Phi^{(0)},
\ldots , \pa_x^{k\cdot r}\Phi^{(0)} \rrb}{W \llb \Phi^{(0)},
\pa_x^{r} \Phi^{(0)}, \ldots , \pa_x^{(k-1)\cdot r}\Phi^{(0)} \rrb}
\lab{sol-10} \\
{1\over r} u_{r-2}^{(k)} \eq Res\, L^{{1\over r}} = \pa_x^2 \ln \t^{(k)} \quad
,
\quad  \t^{(k)} = W \llb \Phi^{(0)}, \pa_x^{r} \Phi^{(0)},
\ldots , \pa_x^{(k-1)\cdot r}\Phi^{(0)} \rrb  \lab{sol-11}
\er
where $\Phi^{(0)}$ is again given explicitly by \rf{sol-6}.
\lskip
{\large {\bf 5. Relation to Constrained Generalized Toda Lattices}}
\lskip
Here we shall establish the equivalence between the set of successive DB
transformations of the $SL(r+1,1)$ KP-KdV system \rf{sol-9} :
\br
L^{(k+1)} &=&  T^{(k)}\; L^{(k)} \; \(T^{(k)}\)^{-1} \qquad ,\quad
T^{(k)} = \Phi^{(k)}  D {\Phi^{(k)} }^{-1}       \lab{seq-a} \\
L^{(0)} &=& D^r + \sum_{l=0}^{r-2} u_l^{(0)} D^l +
\Phi^{(0)} D^{-1} \Psi^{(0)}    \lab{seq-bb}
\er
and the equations of motion of a {\em constrained}
generalized Toda lattice system, underlying the two-matrix string model,
which contains, in particular, the {\em two-dimensional} Toda lattice
equations.

For simplicity we shall illustrate the above property
on the simplest nontrivial case of $SL(3,1)$ KP-KdV hierarchy
\rf{sl-31}.
We note that eqs.\rf{sol-3}--\rf{sol-5} (or \rf{sol-3-aa}--\rf{sol-4-b}) can
be cast in the following recurrence form:
\br
\pa_x \ln A^{(n-1)}& =& B^{(n)} - B^{(n-1)}  \lab{Tod-1} \\
u^{(n)} - u^{(n-1)}& =& 2\pa_x B^{(n)}  \lab{Tod-2} \\
A^{(n)} - A^{(n-1)} &=& \pa_x \( \( B^{(n)}\)^2 + \h \( u^{(n)} + u^{(n-1)}\)\)
\lab{Tod-3}
\er
with ``initial'' conditions (cf. \rf{sol-6}) :
\be
A^{(0)}=B^{(0)}=u^{(0)}=0 \qquad , \quad B^{(1)} \equiv \pa_x \ln \Phi
\lab{Tod-4}
\ee
where $\Phi$ is so far an arbitrary function. Now, we can view
\rf{Tod-1}--\rf{Tod-3} as a system of lattice equations for the dynamical
variables $A^{(n)},B^{(n)},u^{(n)}$ associated with each lattice site $n$
and subject to the boundary conditions:
\be
A^{(n)}=B^{(n)}=u^{(n)}=0 \quad ,\;\; n \leq 0  \lab{bound-cond}
\ee
Taking \rf{Tod-4} as initial data, one can solve the lattice system
\rf{Tod-1}--\rf{Tod-3} step by step (for $n=1,2,\ldots$) and the
solution has precisely the form of \rf{sol-3-aa}--\rf{sol-4-b}.

The lattice system \rf{Tod-1}--\rf{Tod-3} can be identified with the
${\ti t}_1 \equiv x$ evolution equations of the
{\em constrained} generalized Toda lattice hierarchy defined as follows
\ct{BX,enjoy} :
\br
\partder{}{t_r} Q = \llb Q^r_{(+)} , Q \rrb  \quad , \quad
\partder{}{t_r} {\bar Q} = \llb Q^r_{(+)} , {\bar Q} \rrb \quad , \quad
r=1,\ldots , p_1  \lab{L-3} \\
\partder{}{{\ti t}_s} Q = \llb Q , {\bar Q}^s_{-} \rrb     \quad , \quad
\partder{}{{\ti t}_s} {\bar Q} = \llb {\bar Q} , {\bar Q}^s_{-} \rrb
\quad ,\quad s=1,\ldots ,p_2    \lab{L-4}\\
- g \llb Q , {\bar Q} \rrb = \one   \phanta   \lab{string-eq}
\er
Here $Q$ and ${\bar Q}$ are semi-infinite matrices, {\sl i.e.}, with indices
running from $0$ to $\infty$, with the following explicit parametrization:
\br
Q_{nn} = a_0 (n) \quad , \quad Q_{n,n+1} =1 \quad ,\quad
Q_{n,n-k} = a_k (n) \quad k=1,\ldots , p_2 -1   \nonu  \\
Q_{nm} = 0 \quad {\rm for} \;\;\; m-n \geq 2 \;\; ,\;\; n-m \geq p_2 \phanta
\lab{param-1}  \\
{\bar Q}_{nn} = b_0 (n) \quad , \quad {\bar Q}_{n,n-1} = R_n \quad , \quad
{\bar Q}_{n,n+k} = b_k (n) R_{n+1}^{-1} \cdots R_{n+k}^{-1}
\quad k=1,\ldots ,p_1 -1    \nonu  \\
{\bar Q}_{nm} = 0 \quad {\rm for} \;\;\; n-m \geq 2 \;\; ,\;\; m-n \geq p_1
\phanta    \lab{param-2}
\er
The subscripts $-/+$ in \rf{L-3},\rf{L-4} denote lower/upper triangular parts,
whereas $(+)/(-)$ denote upper/lower triangular plus diagonal parts. In the
case under consideration the number $p_2 =3$ in \rf{param-1}, whereas the
number $p_1$ in \rf{param-2} is arbitrary finite or $\infty$ \foot{Both
numbers $p_{1,2}$ indicating the number of non-zero diagonals, outside the
main one, of the matrices ${\bar Q}$ and $Q$ are related with the polynomial
orders of the corresponding string two-matrix model potentials, whereas the
constant $g$ in \rf{string-eq} denotes the coupling parameter between the
two random matrices.}.

Note the presence of the non-evolution constraint eq.\rf{string-eq}, which
is called ``string equation''.
The lattice equations for the matrix elements $a_k (n)$ of $Q$
(the first eqs.\rf{L-3} and \rf{L-4}) can be solved explicitly as functionals
of the matrix elements of ${\bar Q}$ :
\be
Q_{(-)} = \sum_{s=0}^{p_2 -1} \a_s {\bar Q}^s_{(-)}  \qquad ; \quad
\a_s \equiv - (s+1) \frac{{\ti t}_{s+1}}{g}           \lab{2-3}
\ee
Furthermore, it is more convenient to introduce another matrix ${\hat Q}$
(with matrix elements ${\hat R}_n, {\hat b}_k (n)$, cf. \rf{param-2})
in place of ${\bar Q}$ defined as:
\be
{\hat Q}^{p_2 -1}_{(-)} = \sum_{s=0}^{p_2 -1} \a_s {\bar Q}^s_{(-)}  \quad ;
\quad {\hat Q}^{p_2 -1}_{(+)} = I_{+}  \longrightarrow {\hat Q}^{p_2 -1} = Q
\lab{tatko}
\ee
with ${I_{+}}_{nm}= \d_{n+1,m}$, where the last equality follows from \rf{2-3}.
More generally, we have the relations $ {\hat Q}^s_{(-)} = \sum_{\s =0}^s
\g_{s\s} {\bar Q}^{\s}_{(-)}$ for any $s=1,\ldots , p_2$ with coefficients
$\g_{s\s}$ simply expressed through $\a_s$ \rf{2-3}. Specifically we get:
\br
{\hat R}_n &=& \g_{11} R_n \quad , \quad {\hat b}_0 (n) = \g_{11} b_0 (n) +
\frac{\g_{21}}{2\g_{11}} \quad ,\quad {\hat b}_1 (n) = \g_{11}^2 b_1 (n) +
\frac{\g_{31}}{3\g_{11}} - \( \frac{\g_{21}}{2\g_{11}} \)^2  \lab{tatko-1} \\
\g_{11} &\equiv& \( \a_{p_2 -1} \)^{1\over {p_2 -1}} \quad ,\quad
\g_{21} \equiv {2\over {p_2 -1}} \frac{\a_{p_2 -2}}{\(\a_{p_2 -1}\)^{{p_2 -3}
\over {p_2 -1}}}  \nonu \\
\g_{31} &\equiv&{3\over {p_2 -1}} \llb
\frac{\a_{p_2 -3}}{\(\a_{p_2 -1}\)^{{p_2 -4}
\over {p_2 -1}}} - \frac{p_2 -4}{2\( p_2 -1\)}
\frac{\a_{p_2 -2}^2}{\(\a_{p_2 -1}\)^{{2p_2 -5}\over {p_2 -1}}}\rrb
\lab{tatko-2}
\er
Accordingly, the evolution eqs.\rf{L-4} acquire the form:
\be
\partder{}{{\hat t}_s} {\hat Q} = \llb {\hat Q} , {\hat Q}^s_{-} \rrb \qquad
{\rm with} \quad \partder{}{{\hat t}_s} = \sum_{\s =1}^s \g_{s\s}
\partder{}{{\ti t}_{\s}}  \lab{L-4-a}
\ee

The remaining independent lattice equations then read (we write down explicitly
only the ${\hat t}_1 \equiv x$ and ${\hat t}_2$
evolution equations for the $p_2 =3$ case) :
\br
\pa_x \ln {\hat R}_{n+1} \eq {\hat b}_0 (n+1) - {\hat b}_0 (n) \quad ,\quad
{\hat b}_1 (n) - {\hat b}_1 (n-1) = \pa_x {\hat b}_0 (n)  \lab{x-motion-1} \\
{\hat R}_{n+1} - {\hat R}_n \eq \pa_x \( {\hat b}_0^2 (n) + {\hat b}_1 (n) +
{\hat b}_1 (n-1) \)    \lab{x-motion-2} \\
\partder{}{{\hat t}_2} {\hat R}_{n+1} \eq\pa_x \Bigl\lb \pa_x {\hat R}_{n+1} +
2{\hat b}_0 (n) {\hat R}_{n+1} \Bigr\rb   \phanta  \lab{1+1-aa} \\
\partder{}{{\hat t}_2} {\hat b}_0 (n) \eq \pa_x \Bigl\lb 2 {\hat b}_1 (n) +
{\hat b}_0^2 (n) - \pa_x {\hat b}_0 (n) \Bigr\rb   \quad ,\quad
\partder{}{{\hat t}_2} {\hat b}_1 (n) =  \pa_x {\hat R}_{n+1}   \lab{1+1-cc}
\er
Now, we observe that the system of Darboux-\Back ~equations for $SL(3,1)$
KP-KdV hierarchy \rf{Tod-1}--\rf{Tod-3} exactly coincides upon identification:
\be
B^{(n)}= {\hat b}_0 (n-1) \quad ,\quad u^{(n)} = 2 {\hat b}_1 (n-1) \quad
,\quad
A^{(n)}= {\hat R}_n  \lab{Tod-5}
\ee
with the $x\equiv {\hat t}_1 $ constrained Toda lattice evolution equations
\rf{x-motion-1}--\rf{x-motion-2}.
Also, the higher Toda lattice evolution parameters
can be identified with the following subset of evolution parameters of the
$SL(p_2 ,1)$ KP-KdV hierarchy \rf{sol-9} \ct{enjoy,office} :
\be
{\hat t}_s \simeq t_s^{KP-KdV} \quad s=2,\ldots ,p_2 \quad ; \quad
t_r \simeq t_{r(p_2 -1)}^{KP-KdV} \quad r=1,\ldots ,p_1   \lab{param}
\ee
the second identification resulting from \rf{tatko}.

In particular, excluding ${\hat b}_0 (n) \equiv B^{(n+1)}$ and
${\hat b}_1 (n) \equiv \h u^{(n+1)}$ in \rf{Tod-3} using
\rf{x-motion-1}--\rf{1+1-cc}, we obtain the two-dimensional Toda lattice
equation for $A^{(n)}\equiv {\hat R}_n$ :
\be
\pa_x \pa_{{\hat t}_2} \ln A^{(n)} = A^{(n+1)}  - 2 A^{(n)} + A^{(n-1)}
\lab{motion-00}
\ee
\lskip
{\large {\bf 6. Discussion and Outlook}}
\mskp
{\bf 6.1 Partition Function of the Two-Matrix String Model}

The partition function $Z_N$ of the two-matrix string model is simply
expressed in terms of the ${\bar Q}$ matrix element $b_1 (N-1)$ at
the Toda lattice site $N-1$, where $N$ indicates the size ($N \times N$) of
the pertinent random matrices: $\pa_x^2 \ln Z_N = b_1 (N-1)$
(cf. \ct{BX,enjoy}). Thus, using \rf{sol-11} and \rf{sol-6} together with
\rf{param}, and accounting for the relations \rf{tatko-1}--\rf{tatko-2},
 we obtain the following exact
solution at {\em finite} $N$ for the two-matrix model partition function:
\br
Z_N &= & W \llb {\hat \Phi}^{(0)}, \pa_{{\hat t}_{p_2 -1}} {\hat \Phi}^{(0)},
\ldots , \pa_{{\hat t}_{p_2 -1}}^{N-1} {\hat \Phi}^{(0)} \rrb\,
\exp - N \int^{{\hat t}_1}   \( \frac{\g_{21}}{2\g_{11}}\)
\lab{sol-3-matr}  \\
{\hat \Phi}^{(0)} &=& \int_{\Gamma} \frac{d\l}{2\pi} c(\l )
e^{{\hat \xi} \( \l ,\{ {\hat t},t\}\)} \quad ,\;\;
{\hat \xi}\( \l ,\{ {\hat t},t\}\) \equiv \sum_{s=1}^{p_2} \l^s {\hat t}_s
+ \sum_{r=2}^{p_1} \l^{r(p_2 -1)} t_r     \lab{sol-6-matr}
\er
where $x \equiv {\hat t}_1$ and the $\g$-coefficients are defined in
\rf{tatko-2}. The ``density'' function $c(\l )$ in \rf{sol-6-matr} is
determined from matching the expression for ${\hat \Phi}^{(0)}$:
$\pa_x \ln {\hat \Phi}^{(0)} = {\hat b}_0 (0) = \g_{11} b_0 (0) +
\g_{21}/2\g_{11}$ (cf. \rf{Tod-4},\rf{Tod-5},\rf{tatko-1}), with the expression
for $b_0 (0)$ in the orthogonal-polynomial formalism \ct{BX}:
\br
&&\int_{\Gamma} \frac{d\l}{2\pi} c(\l ) \exp \( \sum_{s=1}^{p_2}
\l^s {\hat t}_s + \sum_{r=2}^{p_1} \l^{r(p_2 -1)} t_r \) \nonu \\
\eq \exp \( \int^{{\ti t}_1} \frac{\g_{21}}{2 \g_{11}^2}\)
 \int_\Gamma \int_\Gamma d\l_1 d\l_2 \,
\exp \(\sum_{r=1}^{p_1} \l_1^r t_r + \sum_{s=1}^{p_2} \l_2^s {\ti t}_s +
g \l_1 \l_2 \)\Bgv_{t_1 = {\hat t}_{p_2 -1} ({\ti t})}  \lab{tatko-3}
\er

Obviously, the most important question now is to study the physical {\em
double-scaling} limit \ct{double-scale} of \rf{sol-3-matr}, which amounts to
a special fine-tuned limit $ N \to \infty$. The latter presumably includes
renormalizations and critical point approaching of the $\(t_r ,{\ti t}_s \)$
parameters.
\mskp
{\bf 6.2 Connection to Grassmannian manifolds and
n-Soliton Solution for the KP Hierarchy}.

Let $\{ \psi_1, \ldots, \psi_n \}$ be a basis of solutions of the $n$-th
order equation $L \psi = 0$,
where $L=\( D + v_n \) \( D + v_{n-1} \) \ldots \( D + v_1 \) $.
If $W_k$ denotes the Wronskian determinant of $\{ \psi_1, \ldots, \psi_k\}$
then one can show that \ct{wilson,ince}:
\be
v_i = \pa \( \ln { W_{i-1} \over W_{i}} \) \qquad W_0 =1
\lab{wil}
\ee
This allows to show that the space of differential operators
is parametrized by the Grassmannian manifold (see e.g. \ct{wilson,matsuo}).
Start namely with the given differential operator
$L_n= D^n + u_1 D^{n-1} + \cdots + u_n$ and
determine the kernel of $L_n$ given by $n$-dimensional subspace
of some Hilbert space of functions $\cH$, spanned, let say, by
$\{ \psi_1, \ldots, \psi_n \}$. This establishes the connection one way.
On the other hand let $\{ \psi_1, \ldots, \psi_n \}$ be a basis of one point
$\O$ of ${\rm Gr}^{(n)}$ being a Grassmannian manifold.
Define the differential equation as $L_n (\O) f =  W_{k} (f) / W_k$.
{}From \rf{kw} this associates the differential operator
\be
L_n= D^n + u_1 D^{n-1} + \cdots + u_n=\( D + v_n \) \( D + v_{n-1} \)
\cdots \( D + v_1 \)
\lab{miura}
\ee
given by a Miura correspondence to a given point of the Grassmannian.

Recall now correspondence (equivalence) between
$N$-generalized two-boson KP and $2N$-boson KP systems ---
eqs.\rf{iss-8aa}--\rf{iss-8c}. The relation \rf{iss-8c} has the form like
in \rf{kw} and, therefore:
\be
\( \llb \( D - B^{(N)}_i \)^{-1} \( D - B^{(N)}_{i+1}\)^{-1}
\cdots \( D - B^{(N)}_N \)^{-1} \rrb^{-1} \)^{\dag} \; \Psi_j = 0
\quad ;\quad i \leq j \leq N
\lab{air}
\ee
The above relations generalize the relations encountered in the study of
flags manifolds and clearly deserve more investigations.

Let us now comment on connection to n-soliton solution for the KP hierarchy.
Assume that the above functions $\psi_i\; i =1 , \ldots, n$ have the
property $\pa_m \psi_i = \pa^m \psi_i $ for arbitrary $m \geq 1$
{}~($\pa_m \equiv \partder{}{t_m}$),
in other words $\psi_i$ are eigenfunctions of $L^{(0)}=D$.
We introduce $ L \equiv L_n D L_n^{-1}$, where $L_n$ is defined in terms
of $\{ \psi_1, \ldots, \psi_n \}$ as in \rf{wil} and \rf{miura}.
It is known that such a Lax operator satisfies a generalized Lax equation
$\pa_m L = \lb L^m_{+}, L \rb$ \ct{Zakh-Dickey,ohta}.

Using \rf{iw} we can rewrite the above Lax operator as a result of successive
DB transformations applied on $D$:
\be
L = L_n D L_n^{-1} = T_n \, T_{n-1}\, \cdots\, T_1 \, D
T_1^{-1} \cdots\, T_{n-1}^{-1} \, T_n^{-1}
\lab{nsolax}
\ee
where $T_i$ are given in terms of Wronskians as in \rf{transf}.
It follows that $L$ can be cast in a form of the Lax operator
belonging to the $n$-generalized two-boson KP hierarchy having the form as in
\rf{f-5} with $r=1, N=n$.
Using the formalism developed in this paper one can prove by induction that
the corresponding $\tau$-function of $L$ takes a Wronskian form
$ \tau_n = W_n \lb \psi_1, \ldots, \psi_n \rb$ reproducing n-soliton solution
to the KP equation derived in \ct{nimmo}.
In fact, choosing $\psi_i = \exp \( \sum t_k \a^k_i\) +
\exp \( \sum t_k \b^k_i\)$
allows to rewrite $\tau_n $ in the conventional form of the
$n$-soliton solution to the KP equation \ct{hirota}.
\lskip
\small
{\bf Acknowledgements.}
E.N. and S.P. gratefully acknowledge support from the Ben-Gurion University,
Beer-Sheva. Also, they extend their sincere gratitude to Prof. K. Pohlmeyer
for cordial hospitality at the University of Freiburg where a major part of
this work was done. S.P. thankfully appreciates financial support by the
{\sl Deutscher Akademischer Austauschdienst} for her visit at the University
of Freiburg.
\mskp
{\bf Note Added.}
After completion of this paper we became aware of refs.\ct{shaw}
where Wronskian expressions for partition functions of matrix models have been
obtained by a different method. We would like to stress that our result
\rf{sol-3-matr}--\rf{sol-6-matr},\rf{tatko-3} explicitly incorporates the
``string-equation'' constraint \rf{string-eq} on the Toda lattice system.

\end{document}